\documentclass[aps,twocolumn,showpacs,preprintnumbers,amsmath,amssymb]{revtex4}
\usepackage{graphicx}
\usepackage{dcolumn}
\usepackage{bm}
\usepackage{dsfont}
\usepackage{psfrag}
\newcommand{\ket}[1]{|\, #1 \,\rangle}
\newcommand{\bra}[1]{\langle\, #1 \,|}
\newcommand{\Tr}{{\rm Tr}}
\newcommand{\uop}{\mathds{1}}
\newcommand{\calP}{{\mathcal P}}
\newcommand{\calL}{{\mathcal L}}

\newcommand{\calU}{{\,{\mathcal U}}}

\begin{document}

\title{Inelastic scattering of light by a cold trapped atom: \\
Effects of the quantum center-of-mass motion}

\author{Marc Bienert}
\affiliation{
Centro de Ciencias F{\'i}sicas UNAM, Campus Morelos UNAM, 62251 Cuernavaca, Morelos, Mexico
}

\author{Wolfgang Merkel}
\affiliation{
Abteilung f\"ur Quantenphysik, Universit\"at Ulm, Albert-Einstein-Allee 11, 89069 Ulm, Germany
}

\author{Giovanna Morigi}
\affiliation{
 Grup d'Optica, Departament de Fisica, Universitat Autonoma de Barcelona, 08193 Bellaterra, Spain
}

\date{\today}

\begin{abstract}
The light scattered by a cold trapped ion, which is in the
stationary state of laser cooling, presents features due
to the mechanical effects of atom-photon interaction. These
features appear as additional peaks (sidebands) in the spectrum of
resonance fluorescence. Among these sidebands the literature has
discussed the Stokes and anti-Stokes components, namely the
sidebands of the elastic peak. In this manuscript we show that the
motion also gives rise to sidebands of the inelastic peaks.
These are not always visible, but, as we show, can be measured in
parameter regimes which are experimentally accessible.
\end{abstract}

 \maketitle

\section{Introduction}

The mechanical effects of photon-atom interactions are at the
basis of several techniques of manipulation of the quantum state
of atoms~\cite{Leibfried03}. Their understanding is therefore
relevant for implementations of coherent control. Features due to the center-of-mass motion
have been measured in the spectrum of resonance
fluorescence of trapped atoms~\cite{Jessen93} and 
ions~\cite{Hoffges97,Raab2000,Tamm2000,Bushev03,Bushev04}. These measurements
allowed to gain insight into the state of the laser-cooled atoms,
and were at the basis of the implementation of cooling schemes on the atomic motion based
on feedback control~\cite{Bushev}.

The spectrum of resonance fluorescence of a laser-cooled atom 
is obtained from the spectral analysis of the intensity of the scattered light. 
By means of these scattering processes the atomic center-of-mass motion is cooled by transferring
center-of-mass energy into the photonic reservoir. Hence, the
scattered photons carry the information about the dynamics that the atom undergoes, which can be partly reconstructed in the power 
spectrum. Most studies done so far on the resonance
fluorescence of trapped ions focused onto the elastically
scattered radiation by a confined atomic dipole and on its Stokes
and anti-Stokes sidebands, thereby finding good agreement between
theory and
experiment~\cite{Lindberg86,Cirac93,Jessen93,Plenio95,Hoffges97,Raab2000,Tamm2000,Bushev03,Bushev04,Bushev}.
On the other hand, the effect of the center-of-mass motion on the
inelastic part of the spectrum of resonance fluorescence are
largely unexplored. Features of mechanical effects in this
radiation, which has no classical analog, have been studied in~\cite{Bienert04} for
the case of a trapped atom, undergoing laser cooling by driving
two coupled dipole transitions in the regime where fluorescence is
solely due to the mechanical effects of light. In~\cite{Bienert04} 
it was found that the quantum motion gives rise to
sidebands of the inelastic spectrum, which can be mapped to Raman
processes between the dressed states of the bare three-level
transition.

In this work we investigate the spectrum of resonance fluorescence
of a trapped atom whose dipole transition is driven by the cooling
laser, and we address in particular the issue whether and how
features due to the mechanical effects of light can be identified
and measured in the inelastic part of the spectrum. Indeed, we
show that the quantum motion gives rise to sidebands of the peaks
of the inelastic spectrum, which have different properties
compared to the sidebands of the elastic component. These sidebands are
not always visible, however they can be measured in experimentally
accessible parameter regimes, like for instance in~\cite{Indium}.

The investigation in this work complements the theoretical study
of~\cite{Cirac93}, which focussed onto the Stokes and anti-Stokes
components of the spectrum of resonance fluorescence of a driven
dipole. Like in~\cite{Cirac93}, we consider the situations when
the cooling laser is (i) a running wave and (ii) a standing wave.
The spectrum is calculated by using the spectral decomposition of
the atomic master
equation~\cite{Barnett2000,Briegel93,Jakob03}

This work is organized as follows. In Sec.~\ref{Sec:Results} we
present and discuss the spectra of resonance fluorescence. In
Sec.~\ref{Sec:Evaluate} the theoretical model at the basis of the
derivation is reported. The conclusions are drawn in
Sec.~\ref{Sec:Conclude}. The appendices report further details of
the derivations of Sec.~\ref{Sec:Evaluate}.

\section{Spectrum of resonance fluorescence}
\label{Sec:Results}

\begin{figure}
\centerline{\includegraphics[width=7cm]{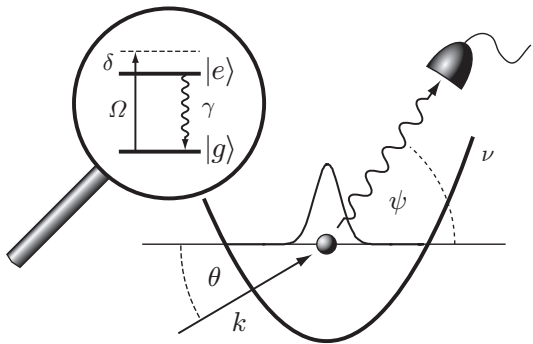}}
\caption{\label{fig:1} An atom is trapped in a harmonic potential
of frequency $\nu$ and its dipole transition is driven by a laser
field with wave vector $k$ and propagating at angle $\theta$
with the axis of motion. The laser cools the atomic motion and a
narrow-band detector records the spectrum of the intensity of the
light which is scattered at angle $\psi$. The Gaussian-function
denotes the center-of-mass wavepacket, whose finite size affects
the properties of the scattered light. The inset shows the
relevant internal atomic structure, composed by the ground and
excited states $\ket{g}$ and $\ket{e}$, which decays at rate
$\gamma$ into $\ket{g}$. Here, $\Omega$ and $\delta$ denote the laser
Rabi-frequency and detuning, respectively.  }
\end{figure}

\begin{figure*}
\centerline{\includegraphics[width=14cm]{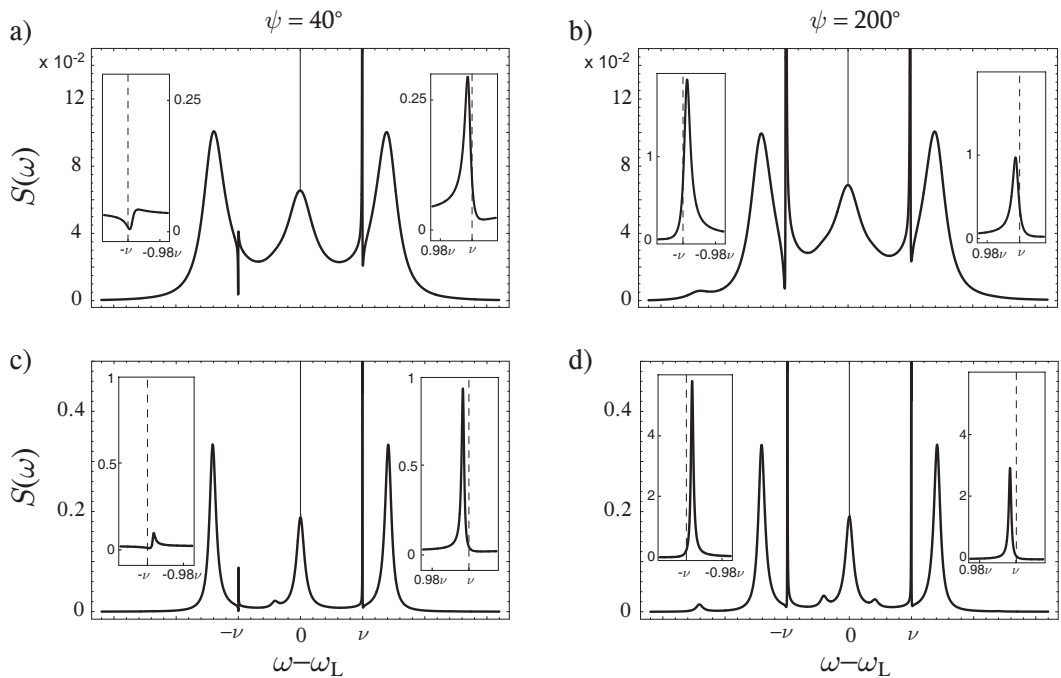}}
\caption{\label{fig:tw} Spectrum of the intensity (in arbitrary units) of the light
scattered by the dipole transition of a trapped ion in the
Lamb-Dicke regime. The ion is sideband-cooled by a traveling wave
laser.
The parameters are $\delta=-\nu$, $\Omega = \nu$,
$\eta=0.1$ and (a)-(b) $\gamma = 0.33\nu$, corresponding to
average phononic occupation $\langle n\rangle=0.15$; (c)-(d)
$\gamma = 0.1\nu$ ($\langle n\rangle=0.14$). Spectra (a) and (c)
are measured at the detection angle $\psi=40^\circ$; Spectra (b)
and (d) at $\psi=200^\circ$. The plots show the elastic peak,
symbolized by the thin line at frequency $\omega=\omega_L$, the
Mollow-spectrum and the motional sidebands of the Mollow peaks.
The thin and outstanding peaks at about $\omega_L\pm \nu$ are the
sidebands of the elastic peak. A magnification of these
contributions is shown in the insets.}
\end{figure*}

Let us consider the radiation scattered by a trapped atom in the
latest stages of laser cooling and detected at a narrow-band
detector, in the setup sketched in Fig.~\ref{fig:1}. The laser
drives the atomic dipole transition, and the atomic motion is laser-cooled to a
stationary state, which is determined by the statistical
equilibrium between scattering processes changing the motional
excitations. The scattered photons are originated from these
processes, and the correlation functions and corresponding spectra
contain some features of these dynamics. The spectrum of the
intensity of the scattered radiation at the detector
is~\cite{Lindberg86}
\begin{eqnarray}
S(\omega)= \chi{\rm Re}\int_0^{\infty}{\rm d}\tau{\rm e}^{-{\rm
i}\omega\tau} \langle E^{(+)}(t+\tau)E^{(-)}(t)\rangle
\end{eqnarray}
where $\chi$ is a constant, $E^{(+)}(t)$ ($E^{(-)}(t)$) are the
negative (positive) frequency part of the electric field at the
detector at time $t$, and $\langle\cdot\rangle$ describes the
average over the atom and electromagnetic field degrees of
freedom. For suitably chosen detection angles the field at the
detector is the field scattered by the source, and it is thus
determined by the state of the atom. 

Let us now make some preliminary considerations, and discuss in
particular the form of the spectrum when the features due to the mechanical
effects are negligible. In this regime, the spectrum of resonance fluorescence of
a dipole transition is constituted by two contributions: The
elastic component, namely a narrow peak at the frequency $\omega_L$ of
the laser, corresponding to Rayleigh scattering, and the inelastic
component, which originates from quantum fluctuations of the atomic
dipole. This part exhibits the well-known peaks of the Mollow
triplet which is found when the transition is driven at
saturation~\cite{AtomPhoton,Mollow}.

The mechanical effects of light clearly modify the form of this
spectrum. The atomic motion, confined by a harmonic potential of
frequency $\nu$, couples to the scattered light. This gives rise to
sidebands of the elastic peak, whose origin can be intuitively understood
on the basis of a classical model for the atomic motion~\cite{Stenholm86}. In the regime
where the atomic wave packet is localized over the wave length of
light, the mechanical coupling is small and scales with the
Lamb-Dicke parameter $\eta\ll 1$. In this regime only the Stokes
and anti-Stokes sidebands are visible. At low
temperatures the form of these sidebands depends critically
on the detection angle. This dependence is due to interference between
quantum paths of excitations coupling internal and external
degrees of freedom~\cite{Cirac93}. 

The inelastic component of the spectrum of resonance fluorescence has no
classical analog. The mechanical effects on this part mainly consists in the appearance of 
sidebands of the inelastic peaks. As we discuss below, these sidebands share
some properties with but also substantial differences from the
sidebands of the elastic peak.

\subsection{Results}

In this section we report and discuss the theoretical results,
whose derivation is reported in Sec.~\ref{Sec:Evaluate}.
Figure~\ref{fig:tw} displays the spectra of the light scattered by
the dipole transition of an ion which is
sideband-cooled~\cite{Stenholm86,Eschner03}. The spectra have been
evaluated for different parameter regimes. Figures~\ref{fig:tw}(a)
and~(b) correspond to the spectrum at two different detection
angles in the situation in which the linewidth of the dipole
transition $\gamma=\nu/3$. Here, the mechanical effects of
atom-photon interaction are clearly evident in the sidebands of
the elastic peak, the two narrow signals at frequency
$\omega_L\pm\nu$. Their width scales with the recoil frequency and
corresponds with the cooling rate~\cite{Lindberg86,Cirac93,Bienert04}. These resonances are magnified
in the inset, where one can see that their centers are slightly
shifted from the values $\omega_L\pm\nu$. This shift is due to the mechanical
coupling~\cite{Cirac93,Bienert04}. The corresponding curves are a superposition of Lorentz
and Fano-like profiles, whereby one functional dependence can
dominate over the other depending on the angle of observation. The
asymmetry of these peaks is due to interference effects, which
originate from the quantum nature of the ion center-of-mass
motion, and whose magnitude depends on the detection
angle~\cite{Cirac93}. Manifestations of the mechanical effects in
the inelastic spectrum are small but yet visible in
Fig.~\ref{fig:tw}(b), causing the appearance of an additional
resonance on the left part of the curve (at
$\omega=\omega_L-2.4\nu$) and a broadening of the central
inelastic peak. The width of the sidebands of the inelastic peaks
is substantially the same as the one of the corresponding
inelastic peak at zero order in the motion. Their height scales
with the recoil frequency, and, like the sidebands of the elastic
peak, it is a non-trivial function of the average phononic
excitation $\langle n\rangle$ and of the angle of detection. Their
functional form is reported in Sec.~\ref{Sec:Evaluate:1}.

Figures~\ref{fig:tw}(c) and~(d) display the spectra of
fluorescence when the linewidth $\gamma=\nu/10$. Here, the
mechanical effects can be clearly identified in the inelastic
spectrum, displaced by about $\pm \nu$ from the corresponding
signal of the Mollow triplet. The sidebands of the central
inelastic peak are centered at the same frequency as the Stokes
and anti-Stokes signals, and give rise to a small broadening at
their basis. 

\begin{figure}
\centerline{\includegraphics[width=8cm]{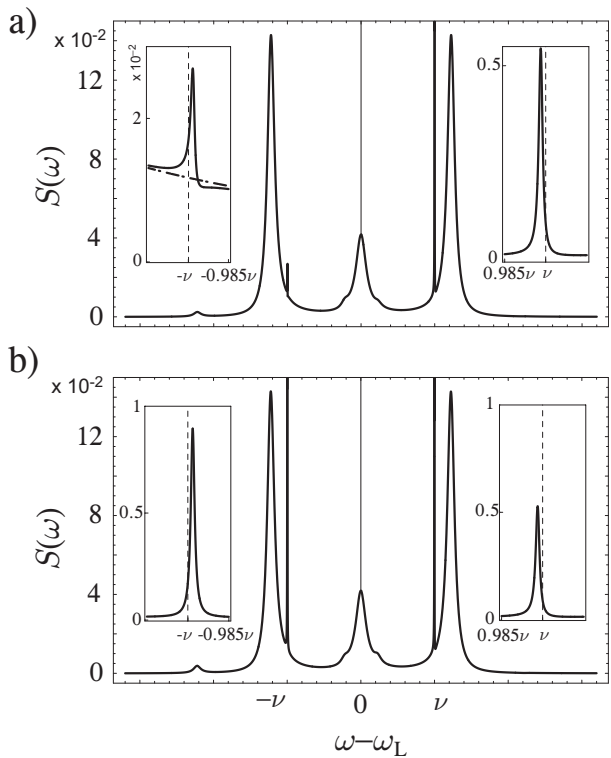}}
\caption{\label{fig:sw} Spectrum of the intensity (in arbitrary units) of the light
scattered by the dipole transition of a trapped ion in the
Lamb-Dicke regime. The ion is sideband-cooled by a standing wave
laser. The parameters are $\delta=-\nu$, $\Omega = \nu$, $\gamma =
0.1\nu$, $\eta=0.1$ and detector angle (a) $\psi=40^\circ$; (b)
$\psi=200^\circ$. The standing wave is described by the function
$\cos(kx+\varphi)$ where $\varphi=\pi/4$ is the value at the trap
center. The mean phononic excitation is $\langle n\rangle =0.04$.
The insets display the sidebands of the elastic peak, the
dashed-dotted lines shows the inelastic spectrum contribution.}
\end{figure}

\begin{figure}
\centerline{\includegraphics[width=8cm]{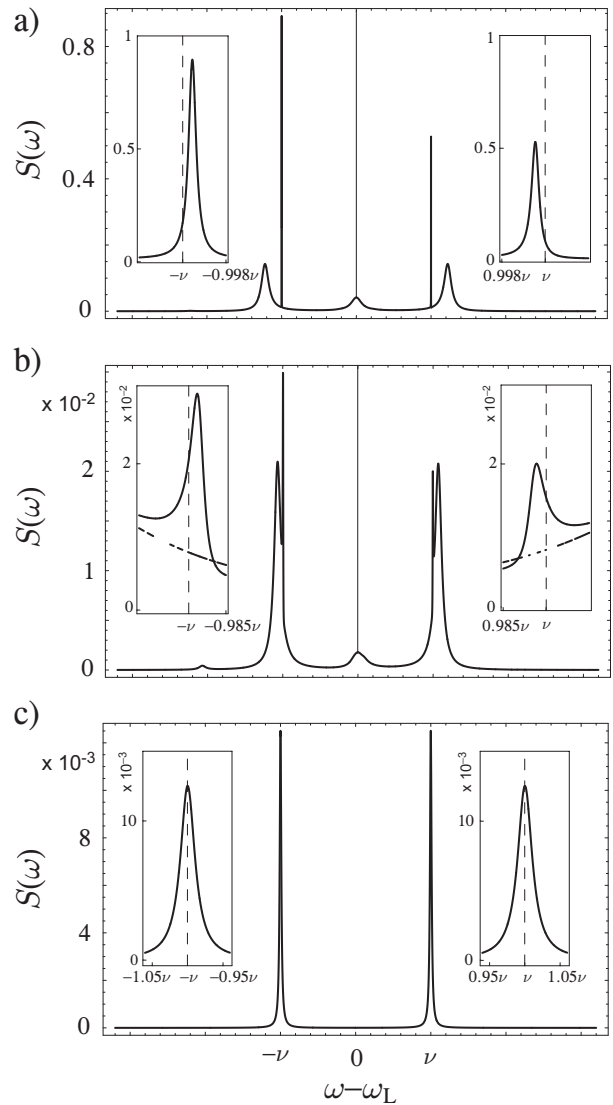}}
\caption{\label{fig:swnode} Same as Fig.~\ref{fig:sw} for
different position in the standing wave. Here, (a) $\varphi =
\pi/4$ ($\langle n\rangle=0.04$); (b) $\varphi=3\pi/8$ ($\langle
n\rangle=0.002$); (c) $\varphi=\pi/2$ ($\langle n\rangle=0.0006$). The
parameters are $\delta=-\nu$, $\Omega = \nu$, $\gamma =0.1\nu$,
$\eta=0.05$, $\psi = 200^\circ$. The insets display the sidebands
of the elastic peak, the dashed-dotted lines correspond to the
inelastic spectrum contribution.}
\end{figure}

We remark that the height of the
mechanical sidebands -- for elastic \emph{and} inelastic
scattering -- critically depends on the detection angle, as it is
visible in the figures. Also for the inelastic spectrum this
effect is caused by interference of the underlying atomic
processes contributing to each spectral signature and depending on
the direction of photon absorption and emission. Thus, the behavior also depends 
on whether the drive is a traveling
wave, as in Fig.~\ref{fig:tw}, or a standing wave.
Figure~\ref{fig:sw} shows the spectrum of an ion where the trap
center is in a point of the standing wave (away from the nodes and
anti-nodes), for $\gamma=\nu/10$ and at two different detection
angles. Here, one can observe the dependence of the sidebands of the
elastic and inelastic part on the detection angle. 
Figure~\ref{fig:swnode} displays the spectra of resonance
fluorescence at various points of the standing wave. For $\varphi
= \pi/4$, Fig.~\ref{fig:swnode}(a), the spectrum is governed by
the outstanding sidebands of the elastic peak, overtopping the
Mollow triplet. At $\varphi=3\pi/8$, Fig.~\ref{fig:swnode}(b), the
trap is closer to the node of the standing wave. Here, the
resonances of the Mollow triplet get closer. In this case, the
height of the sidebands of the elastic peak decrease and the
sidebands of inelastically scattered light are visible. A peculiar
case is found when the atom is placed at the node,
Fig.~\ref{fig:swnode}(c) at $\varphi = \pi/2$. Here, the only
spectral signals are the motional sidebands. These sidebands are now perfect
Lorentzian curves of identical shape independent of the detector
position, as already pointed out in~\cite{Cirac93}.
What is novel is that all other contributions of the spectrum,
elastic and inelastic, disappear due to the vanishing of the
driving field at the center of the trap. In particular, even the
motional sidebands of the inelastic spectrum disappear. 
The appearance of the sidebands of the elastic peak is due to 
the effects of the gradient of
the field intensity over the atomic wavepacket at the node.
The suppression of inelastic scattering processes, included the ones that change the
motional excitation, can be understood by
considering that there are no quantum fluctuations of the atomic
dipole at the node of a standing wave.

These behaviors are discussed in detail in the following
theoretical treatment. It should be remarked that the sidebands of the inelastic component
become visible when the transition, at zero order in the mechanical
effects, is driven at saturation. Below saturation the mechanical
effects in the inelastic spectrum manifest in small sidebands of
the Lorentz curve centered at the laser frequency. These sidebands
are at the same frequencies as the ones of the Stokes and
anti-Stokes components of the elastic peak, and are covered by
these signals.

\section{Evaluation of the spectrum of resonance fluorescence}
\label{Sec:Evaluate}

We consider an atom of mass $M$, whose center of mass motion is
confined by a harmonic potential of frequency $\nu$. 
We assume that the relevant motion
is along one dimension, the $x$-axis, while the radial motion is
frozen out. A laser, of frequency $\omega_L$ and wave vector $k$,
couples quasi-resonantly with the atomic dipole transition between
the ground and excited states $|g\rangle$ and $|e\rangle$ at
frequency $\omega_0$ and linewidth $\gamma$. The motion is in the Lamb-Dicke regime, namely
the size of the center-of-mass wave packet is assumed to be much
smaller than the laser wavelength, and the dipole linewidth
$\gamma<\nu$. In this regime the laser can be set
at frequency $\omega_L=\omega_0-\nu$ thereby sideband-cooling the
motion~\cite{Stenholm86,Eschner03}. The atomic motion is in the stationary 
state of sideband cooling and scatters the laser light. 

Let a detector measure the radiation scattered at the
angle $\psi$ with the atom motional axis.
In the far-field the scattered field is proportional to
the retarded value of the atomic dipole at the corresponding
position $x$. The spectrum at frequency $\omega$ is given by~\cite{Carmichael93}
\begin{equation}
\label{SomegaPol:2} S(\omega)= \tilde{\chi}(\psi) {\rm
Re}\int_0^{\infty}{\rm d}\tau{\rm e}^{-{\rm
i}(\omega-\omega_{L})\tau} \langle D^\dagger(x,\tau) D(x,0)\rangle_{\rm st}
\end{equation}
where $\tilde{\chi}(\psi)$ is a function of the emission angle and
$D(x,t)$ is dipole lowering operator in the reference frame
rotating with the laser frequency $\omega_{L}$ at the position $x$
of the atomic center of mass. The correlation function in
Eq.~(\ref{SomegaPol:2}) is formally evaluated by means of the
quantum regression theorem, according to which $D(x,t)=D(x) {\rm
e}^{{\cal L} t}$ with ${\cal L}$ the Liouvillian of the master
equation for the atomic density matrix $\varrho$,
\begin{equation}\partial\varrho/\partial t={\cal L}\varrho.
\end{equation}
The average $\langle \cdot \rangle_{\rm st}$ is taken over the atomic
density matrix $\varrho_{\rm st}$ at steady state, solution of the
equation ${\cal L}\varrho_{\rm st}=0$.

The explicit form of $S(\omega)$, Eq.~(\ref{SomegaPol:2}), is
found by applying the spectral decomposition of the Liouville
operator ${\cal L}$ according to the
secular equations
\begin{eqnarray*}
{\cal L}\varrho^{\lambda}&=&\lambda\varrho^{\lambda},\\
\check{\varrho}^{\lambda}{\cal L}&=&\lambda
\check{\varrho}^{\lambda},
\end{eqnarray*}
with eigenvalues $\lambda$ and right and left eigenelements
$\varrho^{\lambda}$ and $\check{\varrho}^{\lambda}$, respectively~\cite{Briegel93,Barnett2000}.
The orthogonality and completeness of the eigenelements is defined
with respect to the trace, such that ${\rm
  Tr}\{\check{\varrho}^{\lambda^{\prime}}\varrho^{\lambda}\} =
\delta^{\lambda^{\prime},\lambda}$. We define the projectors onto
the eigenspace corresponding to the eigenvalue $\lambda$ as ${\cal
P}^{\lambda}=\varrho^{\lambda}\otimes \check\varrho^{\lambda}$
such that their action on an operator $X$ is defined as ${\cal
P}^{\lambda}X=\varrho^{\lambda}{\rm
Tr}\{\check{\varrho}^{\lambda}X\}$, and they thus fulfill ${\cal
L}{\cal P}^{\lambda}={\cal P}^{\lambda}{\cal L}=\lambda{\cal
P}^{\lambda}$. By applying this formalism, we rewrite
Eq.~(\ref{SomegaPol:2}) as
\begin{equation}
\label{S:3} S(\omega)=\tilde{\chi}(\psi){\rm Re} \sum_{\lambda}\frac{1}{{\rm
i}(\omega-\omega_{L})-\lambda} {\rm  Tr}\left\{D^{\dagger}(x)
{\cal P}^{\lambda}D(x)\varrho_{\rm st}\right\}.
\end{equation}
where we have used the completeness relation of the eigenelements
of ${\cal L}$. Although completeness of this kind of basis is
generally not warranted, the spectral decomposition that we
consider in the following is complete, and the spectrum of
resonance fluorescence can be thus cast in the form of
Eq.~(\ref{S:3}).

\subsection{Model}

We introduce now the basic elements determining the dynamics of
the atom.
The coupling with radiation is assumed to be in the Lamb-Dicke
regime. This regime
is characterized by the small value of the Lamb-Dicke parameter
\begin{equation}
\eta=\sqrt{\frac{\hbar k^2}{2M\nu}},
\end{equation}
which scales the coupling between internal and external atomic degrees of freedom
due to photon scattering.
We identify $\eta$ as the parameter of the perturbative expansion.
According to this expansion, at second order the spectrum in
Eq.~(\ref{S:3}) is decomposed into the terms
\begin{equation}
S(\omega)=S_0(\omega)+S_1(\omega)+S_2(\omega)+{\rm O}(\eta^3)
\end{equation}
where the subscripts label the corresponding order in the
Lamb-Dicke expansion. In order to evaluate $S_j(\omega)$,
we consider the operators ${\cal L}$ and $D$ at second order in
the expansion in $\eta$. The dipole $D(x)=D_0+D_1+D_2+{\rm
O}(\eta^2)$, where the individual terms have the form
\begin{eqnarray*}
&&D_0=\sigma\\
&& D_1=-{\rm  i}k x\cos\psi\,\sigma,\\
&&D_2=-\frac{1}{2}k^2x^2\cos^2\psi\,\sigma,
\end{eqnarray*}
with $\sigma=|g\rangle\langle e|$. The Liouville operator is
decomposed into the terms ${\cal L}={\cal L}_0+{\cal L}_1+{\cal
L}_2$, which we now introduce in detail. At zero order internal
and external degrees of freedom are decoupled, namely
\begin{equation}
{\cal L}_0={\cal L}_{\rm I}+{\cal L}_{\rm E}
\end{equation}
where ${\cal L}_{\rm E}$ and ${\cal L}_{\rm I}$ act on the external and
internal degrees of freedom, respectively. Here, \begin{equation}
{\cal L}_{\rm E}\varrho=\frac{1}{{ i}\hbar}[H_{\rm  mec},\varrho]
\end{equation}
with
\begin{equation}
H_{\rm  mec}= \hbar\nu\left(a^{\dagger}a+\frac{1}{2}\right),
\end{equation}
where $a$ and $a^{\dagger}$ are the annihilation, creation
operators of a quantum of energy $\hbar\nu$, respectively. We
denote with $|n\rangle$ the eigenvectors of $H_{\rm mec}$,
fulfilling $H_{\rm mec}|n\rangle=\hbar \nu (n+1/2)|n\rangle$ with
$n=0,1,2,\ldots$ The term ${\cal L}_{\rm I}$ acts on the internal
degrees of freedom and it is defined as
\begin{eqnarray}
{\cal L}_{\rm I}\varrho=\frac{1}{{ i}\hbar}[H_0+V_0,\varrho]+{\cal
K}\varrho
\end{eqnarray}
where $H_0=\hbar\delta|g\rangle\langle g|$, with
$\delta=\omega_{L}-\omega_0$, and
\begin{eqnarray*}
&&V_0=\frac{1}{2}\hbar \Omega \zeta(\varphi)\sigma^{\dagger}+{\rm H.c.}\\
&&{\cal
    K}_{0}\varrho=\frac{\gamma}{2}
\left(2\sigma\varrho\sigma^{\dagger}
    -\sigma^{\dagger}\sigma\varrho-\varrho\sigma^{\dagger}\sigma\right)
\end{eqnarray*}
are the interaction with radiation at zero order in the mechanical
effects. Here, we have denoted by $\Omega \zeta(\varphi)$ the laser Rabi
frequency, whereby $\zeta(\varphi)$ is a dimensionless function of
the angle $\varphi$, which is $\zeta(\varphi)=\exp({\rm
i}\varphi)|_{\varphi=0}$ for a traveling wave drive, and is given by
$\zeta(\varphi)=\cos\varphi$ for a standing wave drive.

The first and second order Liouvillian give rise to the mechanical
coupling between internal and external degrees of freedom,
\begin{eqnarray}
{\cal L}_{1}\varrho &=&\frac{1}{{\rm
    i}\hbar}\left[xV_1,\varrho\right],\label{L:1}\\
{\cal L}_{2}\varrho &=&
    \frac{1}{2{\rm    i}\hbar}\left[x^2V_{2},\varrho\right]+{\cal
    K}_{2}\varrho. \label{L:2}
\end{eqnarray}
where
\begin{eqnarray}
&&V_1=\frac{1}{2}k\cos\theta\Omega\zeta^{\prime}(\varphi)\sigma^{\dagger}+{\rm H.c.}\label{eq:V1}\\
&&V_2=\frac{1}{2}k^2\cos^2\theta\Omega\zeta^{\prime\prime}(\varphi)\sigma^{\dagger}+{\rm H.c.}\\
&&{\cal K}_{2}\rho=\beta \frac{\gamma}{2}
    k^2\sigma\left(2  x\rho x-x^2\rho-\rho
    x^2\right)\sigma^{\dagger}
\end{eqnarray}
Here, $\theta$ is the angle between the laser and the trap axis,
$\beta$ is a constant, which gives the average recoil due to
spontaneous emission of a photon, and $\zeta^{\prime}(\varphi)$,
$\zeta^{\prime\prime}(\varphi)$ are the first and second derivative
of $\zeta(\varphi)$ with respect to $\varphi$.

The spectral decomposition of the Liouville operator is obtained
using the eigenvalues and eigenelements of operator ${\cal L}_0$,
of which we evaluate the corrections at second order in $\eta$. We
denote the eigenvalues and eigenelements of ${\cal L}_0$ by
$\lambda_0$ and $\varrho_{0}^{\lambda}$,
$\check{\varrho}_{0}^{\lambda}$, respectively. Since internal
and external degrees of freedom are decoupled at zero order in
$\eta$, the eigenvalues of ${\cal L}_0$ are
$\lambda_0=\lambda_{\rm I}+\lambda_{\rm E}$, and the eigenelements
\begin{eqnarray*}
\varrho_{0}^{\lambda}=\rho^{\lambda_{\rm I}}\mu^{\lambda_{\rm E}}
\end{eqnarray*}
where ${\cal L}_{\rm I}\rho^{\lambda_{\rm I}}=\lambda_{\rm I}\rho^{\lambda_{\rm I}}$ and
${\cal L}_{\rm E}\mu^{\lambda_{\rm E}}=\lambda_{\rm E}\mu^{\lambda_{\rm E}}$.
Correspondingly, the projector into the subspace at the eigenvalue
$\lambda_0$ is $${\cal
P}^{\lambda}_{0}=\calP^{\lambda_{\rm I}}\calU^{\lambda_{\rm E}}$$ whereby their
action on the operator $X$ is defined as
\begin{eqnarray*}
&&\calP^{\lambda_{\rm I}}X=\rho^{\lambda_{\rm I}}{\rm
Tr}_{\rm I}\{\check{\rho}^{\lambda_{\rm I}}X\}\\
&&\calU^{\lambda_{\rm E}}X=\mu^{\lambda_{\rm E}}{\rm Tr}_{\rm E}\{\check{\mu}^{\lambda_{\rm
E}}X\}
\end{eqnarray*}
and ${\rm Tr}_{\rm I}$ (${\rm Tr}_{\rm E}$) denotes the trace over the
internal (external) degrees of freedom.

The spectrum of ${\cal L}_{\rm I}$ characterizes the dynamics of the
two-level transition and the spectral properties of the radiation
emitted by the bare atom~\cite{Jakob03}. The eigenvalues of ${\cal L}_{\rm E}$ take on
the values $\lambda_{\rm E}={ i}\ell\nu$, with $\ell=0,\pm 1,\pm
2,\ldots$ Each eigenspace at $\lambda_{\rm E}$ is infinitely degenerate,
and the corresponding left and right eigenelements are, for
instance, $\check{\mu}_n^{\ell}=|n+\ell\rangle\langle n|$,
$\mu^{\ell}_n=|n\rangle\langle
    n+\ell|$. These eigenelements constitute a complete and orthonormal
basis over the eigenspace at this eigenvalue. In particular, the
projector over the eigenspace at $\lambda_{\rm E}={ i}\ell\nu$ is
defined on an operator $X$ as
\begin{eqnarray}
\label{PE} \calU^{\lambda_{\rm E}={ i}\ell\nu}X &=&\sum_n \mu^{\ell}_n
{\rm Tr}_{\rm E}
\{\check{\mu}_n^{\ell}X\}\\
&=&\sum_n |n\rangle\langle n|X|n+\ell\rangle\langle
n+\ell|,\nonumber
\end{eqnarray}
where ${\rm Tr}_{\rm E}$ denotes the trace over the external degrees of
freedom.

At higher orders in the expansion in $\eta$, internal and external
degrees of freedom are coupled, and the degeneracy of the
subspaces at eigenvalue $\lambda_{\rm E}$ is lifted
\cite{Lindberg84,Stenholm86}. The perturbative corrections to the
eigenvalues $\lambda_0$, to the eigenelements $\rho^{\lambda}_0$,
$\check{\rho}^{\lambda}_0$, and to the projectors ${\cal
P}^{\lambda}_0$ are found by solving iteratively the secular
equations at the same order in the perturbative expansion, and
are given by
\begin{equation} \label{rho:1}
  (1-{\cal P}^{\lambda}_{0})\varrho_1^{\lambda}=-\frac{1-{\cal P}^{\lambda}_{0}}
  {\lambda_0-{\cal L}_0}(\lambda_1-{\cal L}_1)\varrho_0^{\lambda},
\end{equation}
  where ${\cal P}^{\lambda}_{0}$ is the zero-order projector onto the subspace
  at eigenvalue $\lambda$, ${\cal P}^{\lambda}_{0}=\varrho_0^{\lambda}\otimes
  \check{\varrho}_0^{\lambda}$. Using (\ref{rho:1}) we obtain
  \begin{eqnarray} \label{rho:2} &&(1-{\cal
  P}^{\lambda}_{0})\varrho_2^{\lambda}=-\frac{1-{\cal P}^{\lambda}_{0}}{\lambda_0-{\cal
  L}_0} \\
&&\times \Bigl[-(\lambda_1-{\cal L}_1)\frac{1-{\cal
  P}^{\lambda}_{0}}{\lambda_0-{\cal L}_0}(\lambda_1-{\cal L}_1)
+(\lambda_2-{\cal L}_2)\Bigr]\varrho_0^{\lambda}. \nonumber
\end{eqnarray} Analogously,
  one finds the perturbative corrections to the left eigenelements
  $\check{\varrho}^{\lambda}_{0}$. This in turn allows one to evaluate the perturbative
  corrections to the projectors ${\cal P}^{\lambda}_0$, which are
  immediately found by using the explicit form of $\check{\varrho}^{\lambda}_{j}$,
  $\varrho^{\lambda}_{j}$
  into the relations
\begin{eqnarray}
&&{\cal
P}^{\lambda}_{1}=\varrho^{\lambda}_{0}\check{\varrho}^{\lambda}_{1}
+\varrho^{\lambda}_{1}\check{\varrho}^{\lambda}_{0}\\
&&{\cal
P}^{\lambda}_{2}=\varrho^{\lambda}_{0}\check{\varrho}^{\lambda}_{2}
+\varrho^{\lambda}_{1}\check{\varrho}^{\lambda}_{1}
+\varrho^{\lambda}_{2}\check{\varrho}^{\lambda}_{0}
\end{eqnarray}
The equations for the corrections $\lambda_1$, $\lambda_2$ to
$\lambda_0$ are
  \begin{eqnarray} \label{lambda:1} \lambda_1&=&{\rm
  Tr}\{\check{\varrho}^{\lambda}_{0}{\cal L}_1\varrho^{\lambda}_{0}\}=0,
\\ \lambda_2&=&{\rm
  Tr}\{\check{\varrho}^{\lambda}_{0}{\cal L}_2\rho^{\lambda}_{0}\}+ {\rm
  Tr}\{\check{\rho}^{\lambda}_{0}{\cal
  L}_1\rho^{\lambda}_{1}\}\label{lambda:2},\nonumber\\ &=&{\rm
  Tr}\{\check{\rho}^{\lambda}_{0}{\cal L}_2\rho^{\lambda}_{0}\}
+{\rm Tr}\{\check{\rho}^{\lambda}_{0}{\cal L}_1\frac{1-{\cal
  P}^{\lambda}_{0}} {\lambda_0-{\cal L}_0}{\cal L}_1\rho^{\lambda}_{0}\},
\end{eqnarray}
where we have used relation ${\cal P}^{\lambda}_{0}{\cal
  L}_1 {\cal P}^{\lambda}_{0}=0$ in Eq.~(\ref{lambda:1}).

A relevant eigenelement of this spectral decomposition is the one
at $\lambda_0=0$, namely the steady-state density matrix. At zero
order this is given by $\varrho_0^{\lambda=0}=\rho_{0}\mu$, where $\mu$
is the density matrix for the external degrees of freedom in the
final stage of the laser-cooling dynamics and has the form
\begin{equation}
\label{mu} \mu=\frac{1}{1+\langle
    n\rangle}\left( \frac{\langle n\rangle}{1+\langle
      n\rangle}\right)^{a^{\dagger}a},
\end{equation}
and
\begin{equation}
\langle n\rangle={\rm Tr}\{a^{\dagger}a\mu\}
\end{equation}
is the average phonon number at steady state. Correspondingly,
$\check{\varrho}_{0}^{\lambda=0}=\uop_{\rm I}\uop_{\rm E}$, where $\uop_j$ is the identity for the
Hilbert space of the internal ($j={\rm I}$) and the external ($j={\rm E}$) degrees of freedom.

\subsection{Explicit form of the sidebands of the elastic and inelastic components}
\label{Sec:Evaluate:1}

We now proceed in evaluating the spectrum of resonance
fluorescence using the terms introduced in the previous section.
At zero order in the mechanical effects, the spectrum is
\begin{eqnarray}
S_0(\omega)=\tilde{\chi}(\psi) {\rm Re}\sum_{\lambda_0}\frac{1}{{\rm
i}(\omega-\omega_{\rm L})-\lambda} G(\lambda)
\end{eqnarray}
where $G(\lambda)={\rm Tr}\{D_0^{\dagger}{\cal
P}^{\lambda}_{0}D_0\rho_0\}$,
which corresponds to the spectrum of the bare atomic dipole. It
thus exhibit peaks which are located at frequencies equal to the
real part of the eigenvalues $\lambda_{\rm I}$, whose width is
determined by the imaginary part of $\lambda_{\rm I}$. Note that
${\rm Tr}\{D_0^{\dagger}{\cal
P}^{\lambda}_{0}D_0\rho_0\}={\rm Tr}\{\sigma^{\dagger}\rho_0^{\lambda_{\rm I}}\}{\rm Tr}\{\check\rho_0^{\lambda_{\rm I}} \sigma\rho_0\}.$
For later convenience, we introduce the scalar functions
\begin{eqnarray*}
&&g(\lambda_{\rm I})={\rm Tr}\{\sigma^{\dagger}\rho_0^{\lambda_{\rm I}}\}\nonumber\\
\label{eq:g}
&&{\check g}(\lambda_{\rm I})={\rm Tr}\{\check\rho_0^{\lambda_{\rm I}} \sigma\rho_0\},
\end{eqnarray*}
such that with this notation $G(\lambda)=g(\lambda_{\rm I}){\check g}(\lambda_{\rm I})$.

The term $S_1(\omega)=0$, since the motional steady state $\mu$,
Eq.~(\ref{mu}), is a thermal state, and thus diagonal in the
vibrational number basis~\cite{Bienert04}.

The features due to the mechanical effects thus manifest at second
order in the Lamb-Dicke expansion, and the corresponding term has
the form
\begin{eqnarray}
S_2(\omega)=\tilde{\chi}(\psi) {\rm Re}\sum_{\lambda}\frac{1}{{\rm
i}(\omega-\omega_{\rm L})-\lambda}F(\lambda)
\label{eq:s2}
\end{eqnarray}
where
\begin{eqnarray}
F(\lambda)={\sum_{a+b+c+d=2}} {\rm Tr}\{D_a^{\dagger}{\cal
P}^{\lambda}_{b}D_c\rho_d\}
\end{eqnarray}
The function $F(\lambda)$ can be decomposed into three
contributions, namely
\begin{eqnarray}
\label{F}
F(\lambda_{\rm E},\lambda_{\rm I}) &=&F_0 (\lambda_{\rm I})\delta_{\lambda_{\rm E},0}\\
&+& F_+(\lambda_{\rm I})\delta_{\lambda_{\rm E},{ i}\nu}+
F_-(\lambda_{\rm I})\delta_{\lambda_{\rm E},-{ i}\nu}\nonumber
\end{eqnarray}
This decomposition allows one to identify the effects of the quantum
motion on the features of the spectrum. The term $F_0 (\lambda_{\rm I})$
is a second-order correction to the features of the zero order
spectrum, $S_0(\omega)$. Namely, it gives rise to small corrections to
the curves of the zero-oder spectrum, however it does not affect relevantly its general
form. The terms $F_{\pm}(\lambda_{\rm I})$, on the other
hand, give rise to novel peaks centered at the frequencies ${\rm
Im}\{\lambda_{\rm I}\}\pm \nu$, which can be identified with the sidebands of the peaks
appearing in $S_0(\omega)$. These terms have the form
\begin{widetext}
\begin{alignat}{1}
\label{F:+} F_{\rm +}(\lambda_{\rm I}) = & \eta^2 \cos^2\theta\;
r(\lambda_{\rm I}, i\nu) \Big[\left\{r_*(\lambda_{\rm I},
i\nu)+u(\lambda_{\rm I}, i\nu)\right\}\langle
n\rangle -t(\lambda_{\rm I}, i\nu)\Big]\nonumber\\
 -&\eta^2 \cos\theta\cos\psi\; \Bigl[g(\lambda_{\rm I})\big[\left\{r_*(\lambda_{\rm I}, i\nu)+u(\lambda_{\rm I}, i\nu)\right\}\langle
n\rangle-t(\lambda_{\rm I}, i\nu)\big]+ r(\lambda_{\rm I},
i\nu)\check g(\lambda_{\rm I}) \langle
n\rangle\Bigr]\nonumber\\
+&\eta^2 \cos^2\psi \;g(\lambda_{\rm I}) \check g(\lambda_{\rm
I})\langle n\rangle
\end{alignat}
and
\begin{alignat}{1}
\label{F:-} F_{\rm -}(\lambda_{\rm I})= & \eta^2
\cos^2\theta~r(\lambda_{\rm I}, -i\nu)
\Big[\left\{r_*(\lambda_{\rm I}, -i\nu)+u(\lambda_{\rm I}, -i\nu)\right\}(\langle n\rangle+1) + t(\lambda_{\rm I}, -i\nu)\Big]\nonumber\\
 -&\eta^2 \cos\theta\cos\psi
\Big[g(\lambda_{\rm I})\big[\left\{r_*(\lambda_{\rm I},
-i\nu)+u(\lambda_{\rm I}, -i\nu)\right\}(\langle
n\rangle+1)+t(\lambda_{\rm I}, -i\nu)\big] +r(\lambda_{\rm I},
-i\nu)\check g(\lambda_{\rm I})(\langle
n\rangle+1)\Big]\nonumber\\
+&\eta^2 \cos^2\psi \;g(\lambda_{\rm I})\check g(\lambda_{\rm
I})(\langle n\rangle+1),
\end{alignat}
\end{widetext}
where we have used $\eta=kx_0$ with $x_0=\sqrt{\hbar/2m\nu}$ and
we have introduced the functions
\begin{alignat}{1}
r(\lambda_{\rm I},\lambda_{\rm E})&=\frac{1}{\hbar}{\rm
Tr}_{\rm I}\{D_0^{\dagger}(\lambda_{\rm I}+\lambda_{\rm E}-{\cal
L}_{\rm I})^{-1}[V_1,\rho^{\lambda_{\rm I}}_{0}]\}
\label{r}\\
r_*(\lambda_{\rm I},\lambda_{\rm E})&=\frac{1}{\hbar}{\rm
Tr}_{\rm I}\{\check{\rho}^{\lambda_{\rm I}}_{0}D_0(\lambda_{\rm E}+{\cal
L}_{\rm I})^{-1}[V_1,\rho_0]\}
\label{r*}\\
u(\lambda_{\rm I},\lambda_{\rm E})&=-\frac{1}{\hbar}{\rm
Tr}_{\rm I}\{\check{\rho}^{\lambda_{\rm I}}_{0}[V_1,
(\lambda_{\rm I}+\lambda_{\rm E}-{\cal L}_{\rm I})^{-1}D_0\rho_0]\}\nonumber\\
\label{u}\\
t(\lambda_{\rm I},\lambda_{\rm E})&=\frac{1}{\hbar}{\rm
Tr}_{\rm I}\{\check{\rho}^{\lambda}_{0}D_0(\lambda_{\rm E}+{\cal
L}_{\rm I})^{-1}\rho_0V_1\} \nonumber\\
&-\frac{1}{\hbar}{\rm
Tr}_{\rm I}\{V_1 \check{\rho}^{\lambda}_{0}(\lambda_{\rm E}+\lambda_{\rm I}-{\cal
L}_{\rm I})^{-1}D_0\rho_0\} \label{t}.
\end{alignat}
The basic steps that lead to Eqs.~(\ref{F:+})-(\ref{t}) are reported in
App.~\ref{app:A}. The results are plotted and discussed for some parameter regimes in Sec.~\ref{Sec:Results}.

We now discuss some general properties of the equations for two
cases: When the laser, driving the dipole and cooling the motion,
is a traveling wave and when it is a standing wave.

{\it Traveling wave drive}. The case when the driving laser is a traveling wave is found by setting $\zeta^\prime(\varphi)=i$ at $\varphi=0$ in Eq.~(\ref{eq:V1}). We evaluate the terms Eqs.~(\ref{r})~-~(\ref{t}) using the zero-order eigenelements $\rho^\lambda_0$ and $\check\rho^\lambda_0$ determined by the eigenvalue equations $\calL_0\rho_{\rm I}^\lambda=\lambda_0 \rho_{\rm I}^\lambda$ and $\check\rho^\lambda_{\rm I} \calL_0=\lambda_0 \check\rho^\lambda_{\rm I}$, as in~\cite{Jakob03}. The eigenvalue $\lambda_{\rm I}=0$ gives the contribution of the elastic peak. As we focus on the case of a dipole driven at saturation, the three other -- nonvanishing -- eigenvalues of the zero-order spectrum determine the position and the widths of the Mollow peaks.

We characterize the higher-order contributions to the spectrum by their corresponding eigenvalues at zero order and analyze the single terms composing $S_2(\omega)$, Eq.~(\ref{eq:s2}). As $S_0(\omega)$ never vanishes for traveling wave drives, the term $F_0$ gives rise to small corrections to the elastic peak (component at $\lambda_{\rm I}=0$) and to the Mollow spectrum (components at $\lambda_{\rm I}\neq 0$). The terms $F_{\pm}(\lambda)$ correspond to novel peaks which add to the ones of the zero-order spectrum. At $\lambda_{\rm I}=0$ these terms give rise to the motional sidebands of the elastic peak, centered about $\lambda_{\rm E}=\pm i\nu$. In this case the terms Eqs.~(\ref{r})~-~(\ref{t}) have to be evaluated for $\check\rho_0^{\lambda_{\rm I}}=\uop_{\rm I}$ and $\rho_0^{\lambda_{\rm I}}=\rho_0$. By
applying the cyclic properties of the trace one finds that $u(0, \lambda_{\rm E})=0$ and that $r_\ast(0, \lambda_{\rm E})=r(0, \lambda_{\rm E})^\ast$. Since $\lambda_{\rm I}=0$, the width and shape of these curves is critically determined by the second order correction $\lambda_2$ to the eigenvalue. The explicit evaluation can be found in~\cite{Lindberg84,Bienert04}. The result we find is in agreement with the one reported in~\cite{Cirac93}.

The mechanical effects on the inelastic part of the spectrum manifest themselves in the components at $\lambda_{\rm I}\neq 0$. They give rise to small shifts of the center-frequency ${\rm Im}\{\lambda_{\rm I}\}$ and of the linewidth ${\rm Re}\{\lambda_{\rm I}\}$, and to novel peaks which are centered at the frequencies
${\rm Im}\lambda_{\rm I}\pm \nu$, and which we discuss in the following. These peaks are the sidebands of the corresponding zero-order curves in the Mollow spectrum. Their width is the width of the corresponding zero-oder Mollow peak, namely ${\rm Re} \{\lambda_{\rm I}\}$. The intensity of the particular signals is proportional to $F_\pm(\lambda_{\rm I})$, where generally all the terms (\ref{r})~-~(\ref{t}), $g(\lambda_{\rm I})$ and $\check g(\lambda_{\rm I})$ contribute. Their shape is in general a superposition of Lorentz (real part of  $F_\pm(\lambda_{\rm I})$) and Fano-like (imaginary part of $F_\pm(\lambda_{\rm I})$) curves.
Note that in the plots shown in Sec.~\ref{Sec:Results}, the real part of $F_\pm(\lambda_{\rm I})$ dominates for the visible peaks and thus these sidebands appear as Lorentz curves in the spectrum. Their visibility,
however, depends on the detection angle. Indeed, also for these terms there are interference effects that give rise to an asymmetry between the right and the left sideband of a given inelastic peak. Note that the sidebands of the central inelastic peak are located at the same frequencies of the Stokes and anti-Stokes sidebands. These terms do not vanish, but are usually hidden by the Stokes and anti-Stokes contributions.
Spectra for different parameter regimes are shown in Fig.~\ref{fig:tw}.

{\it Standing wave drive}. The case when the driving laser is a standing wave is found by setting $\zeta(\varphi)=\cos\varphi$ in $V_0$ and $\zeta^{\prime}(\varphi)=-\sin\varphi$ in Eq.~(\ref{eq:V1}). The general characteristic of the terms composing the spectrum is basically the same as for a traveling wave, the difference arises mainly in the dependence of the curves on the scattering angle. The analytical form of the two sidebands, which we obtain by evaluating explicitly $F_{\pm}(0)$ and the linewidth $\lambda_2$ of the resonances, is in agreement with the formulas reported in~\cite{Cirac93}.
A peculiar behavior is found if the atom is placed at the node of the standing wave, $\zeta(\varphi)=0$, which we discuss in the following. Indeed, in the node of a standing wave the electromagnetic field vanishes at zero order in the Lamb-Dicke expansion. Hence, the dipole is excited by processes at higher order in the Lamb-Dicke parameter $\eta$. In other words, in this regime the spectrum of resonance fluorescence is only due to the gradient of the electromagnetic field over the finite size of the ion wave packet, and thus originates from mechanical effects of atom-photon interaction. In the formulas the terms $V_0=V_2=0$ and the steady-state density matrix $\rho_0=|g\rangle\langle g|\mu$. Details over the eigenelements and eigenvalues of ${\cal L}_{\rm I}$ are reported in App.~\ref{app:node}. In this case $D_0\rho_0=0$ and consequently the term $u(\lambda)=0$, see Eqs.~(\ref{u}). For the same reason the second term on the RHS of $t(\lambda)$ vanishes, see Eq.~(\ref{t}). Also the first term on the RHS of $t(\lambda)$ vanishes, as one can verify by taking the explicit form of $V_1$, Eq.~(\ref{eq:V1}), and of the eigenelements. Hence $t(\lambda)=0$. Moreover, in App.~\ref{app:node} we show that  $F_0(\lambda)=0$. The only terms which do not vanish are $r(0,\lambda_{\rm E})$ and $r_{*}(0,\lambda_{\rm E})$ in $F_{\pm}(\lambda)$, Eqs.~(\ref{r})-(\ref{r*}). Therefore, the spectrum is at second order in the Lamb-Dicke expansion, and the only non-vanishing contributions are the Stokes and anti-Stokes sidebands of the elastic peak. Spectra evaluated for different parameters are shown in Figs.~\ref{fig:sw} and \ref{fig:swnode}.

\section{Conclusions}
\label{Sec:Conclude}

The spectrum of resonance fluorescence of a cold trapped ion,
which is in the stationary state of laser cooling,
presents several features due to the mechanical effects of light
on the quantum motion of the atomic center of mass. In this work we
have shown that, in addition to the well-known Stokes and
anti-Stokes components, similar features appear also in the
inelastic spectrum. These features are sidebands of the inelastic
peaks and have no
classical analog. The quantum features in the mechanical effects of atom-photon
interaction become visible at low temperatures, when the motional
state is close to the ground state, and when the transition
linewidth is smaller than the trap frequency, namely in the regime
where the ion can be sideband cooled. This parameter regime can be encountered, 
for instance, in experimental situations like~\cite{Indium}.

In our investigation we have considered the light scattered from a laser
which sideband-cools the motion in two cases: when the drive is a traveling wave 
and when it is a standing wave. A peculiar
behavior is observed when the drive is a standing wave: Here,
depending on the position of the ion in the mode, the spectrum
changes dramatically. In particular, in the node of the standing
wave we find that the spectrum is solely composed by the Stokes
and anti-Stokes sidebands, while any other contribution is at
higher order in the Lamb-Dicke expansion. In this regime we
recover the form of the sidebands components as predicted
in~\cite{Cirac93}. The situation in the node can be compared with a $\Lambda$-system driven at two-photon resonance, where the zero-order contribution vanishes due to the existence of the dark state coherence. In that case, however, sidebands and unshifted contributions of the inelastic spectrum are 
visible, since the electric field does not vanish at zero order in the Lamb-Dicke expansion~\cite{Bienert04}.

In conclusions, we have studied the effects of the quantum
center-of-mass motion on the inelastic part of the spectrum. This
work complements previous theoretical
analysis~\cite{Lindberg86,Cirac93}, which investigated the elastic
part. In the future we will look at how the quantum center-of-mass
motion affects higher order correlation functions of the scattered
light.

\begin{acknowledgments}
M.B. and G.M. acknowledge hospitality from the Centro
Internacional de Ciencia in Cuernavaca, Mexico during the Humboldtkolleg ``Low dimensional systems in quantum optics'', where part of this
work has been done. The authors are grateful to Gloria
Koenigsberger for pivotal support. M.B. is supported by an
Feodor-Lynen fellowship of the Alexander-von-Humboldt foundation. 
W.M. acknowledges financial support by 
the Landesstiftung Baden-W\"{u}rttemberg.
G.M. is supported by the Spanish
Ministerio de Educaccion y Ciencia (Ramon-y-Cajal).
\end{acknowledgments}

\begin{appendix}

\section{Evaluation of $F_{\pm}(\lambda)$}
\label{app:A}

The terms of Eq.~(\ref{F}) which contribute to $F_{\pm}(\lambda)$
are
\begin{eqnarray*}
&&f_1(\lambda)= {\rm Tr}\{D_0^{\dagger}{\cal
P}^{\lambda}_{1}D_0\varrho_1\}+ {\rm Tr}\{D_0^{\dagger}{\cal
P}^{\lambda}_{2}D_0\varrho_0\}\\
&&f_2(\lambda)= {\rm Tr}\{D_1^{\dagger}{\cal
P}^{\lambda}_{0}D_0\varrho_1\}+ {\rm Tr}\{D_1^{\dagger}{\cal
P}^{\lambda}_{1}D_0\varrho_0\}\\
&&f_3(\lambda)= {\rm Tr}\{D_0^{\dagger}{\cal
P}^{\lambda}_{1}D_1\varrho_0\}\\
&&f_4(\lambda)= {\rm Tr}\{D_1^{\dagger}{\cal
P}^{\lambda}_{0}D_1\varrho_0\}
\end{eqnarray*}
We evaluate now their explicit form using the perturbative
expansion using the definitions introduced in Sec.~\ref{Sec:Evaluate}. We denote by the superscript $^{\pm}$ the part of the 
term, which contributes to $F_{\pm}(\lambda)$. The relevant part of the first term can be rewritten as
\begin{alignat*}{1}
f_1^{\pm}&(\lambda) =\frac{k^2 \cos^2{\theta}}{\hbar}r(\lambda_{\rm I},\lambda_{\rm E})\\
& \times\Bigl[{\rm Tr}_{\rm E}\{x \calU^{\lambda_{\rm E}}x\mu\} \Bigl({\rm
Tr}_{\rm I}\{\check{\rho}^{\lambda_{\rm I}}_{0}D_0(\lambda_{\rm E}+{\cal
L}_{\rm I})^{-1}[V_1,\rho_0]\}\\
& -{\rm Tr}_{\rm I}\{\check{\rho}^{\lambda_{\rm I}}_{0}[V_1,
(\lambda_{\rm I}+\lambda_{\rm E}-{\cal L}_{\rm I})^{-1}D_0\rho_0]\}\Bigr)\\  &+{\rm
Tr}_{\rm E}\{x \calU^{\lambda_{\rm E}}[x,\mu]\}\Bigl({\rm
Tr}_{\rm I}\{\check{\rho}^{\lambda_{\rm I}}_{0}D_0(\lambda_{\rm E}+{\cal
L}_{\rm I})^{-1}\rho_0V_1\}\\  &-{\rm
Tr}_{\rm I}\{V_1\check{\rho}^{\lambda_{\rm I}}_{0}(\lambda_{\rm I}+\lambda_{\rm E}-{\cal
L}_{\rm I})^{-1}D_0\rho_0\}\Bigr)\Bigr]
\end{alignat*}
where we have used the function $r(\lambda_{\rm I},\lambda_{\rm E})$ defined
in Eq.~(\ref{r}). Note that for $\lambda_{\rm I}=0$ it reduces to the
function $-r(\lambda_{\rm E})$ in~\cite{Cirac93}. Using
definitions~(\ref{r*}),~(\ref{u}), and~(\ref{t}) we obtain
\begin{alignat*}{1}
f_1^{\pm}(\lambda) &= k^2 \cos^2 \theta \; r(\lambda_{\rm I}+\lambda_{\rm E})\times\\
&\Bigl[\{r_*(\lambda_{\rm I},\lambda_{\rm E})+u(\lambda_{\rm I},\lambda_{\rm E})\}\;{\rm Tr}_{\rm E}\{x\calU^{\lambda_{\rm E}}x\mu\}\\
&+t(\lambda_{\rm I},\lambda_{\rm E})\;{\rm Tr}_{\rm E}\{x\calU^{\lambda_{\rm E}}[x,\mu]\}\Bigr].
\end{alignat*}
It can be simply verified that function $r_*(\lambda_{\rm I},\lambda_{\rm E})$
in Eq.~(\ref{r*}) fulfills the relation
$r_*(0,\lambda_{\rm E})=r(0,\lambda_{\rm E})^*$. Moreover, the term
$u(0,\lambda_{\rm E})=0$. Finally, the function $t(\lambda_{\rm I},\lambda_{\rm E})$,
defined in Eq.~(\ref{r}), reduces for $\lambda_{\rm I}=0$ to the
function $t(\lambda_{\rm E})$ in~\cite{Cirac93}.

Using these definitions, the other terms can be rewritten
as
\begin{alignat*}{1}
f_2^{\pm}(\lambda) &=-k^2 \cos\theta\cos\psi \; g(\lambda_{\rm I})\times\\
&\Bigl[\{r_*(\lambda_{\rm I},\lambda_{\rm E})+u(\lambda_{\rm I},\lambda_{\rm E})\} {\rm Tr}_{\rm E}\{x \calU^{\lambda_{\rm E}}x\mu\}\\
&+t(\lambda_{\rm I},\lambda_{\rm E}){\rm
Tr}_{\rm E}\{x \calU^{\lambda_{\rm E}}[x,\mu]\}\Bigr]
\end{alignat*}
and
\begin{alignat*}{1}
f_3^{\pm}(\lambda) &=-k^2\cos{\theta} \cos{\psi}\; \check  g(\lambda_{\rm I})
r(\lambda_{\rm I},\lambda_{\rm E})\; {\rm Tr}_{\rm E}\{x \calU^{\lambda_{\rm E}}x\mu\}\\
f_4^{\pm}(\lambda) &=k^2\cos^2{\psi}\;g(\lambda_{\rm I})\check g(\lambda_{\rm I})\;
 {\rm Tr}_{\rm E}\{x\calU^{\lambda_{\rm E}}x\mu\}
\end{alignat*}
where we have used the definitions from Eq.~(\ref{eq:g}).
The trace terms over the external degrees of freedom are
conveniently evaluated using the basis set corresponding to the
projectors in Eq.\ (\ref{PE}), giving
\begin{alignat}{1}
 &{\rm
    Tr}_{\rm E}\{x \calU^{\lambda_{\rm E}} x\mu\}
=x_0^2\left[\delta_{{\lambda_{\rm E}},-{ i}\nu}(\langle n
    \rangle+1)+\delta_{{\lambda_{\rm E}},{ i}\nu}\langle n  \rangle\right],\label{eq:etr1}\\
& {\rm
    Tr}_{\rm E}\{x\calU^{\lambda_{\rm E}} [x,\mu]\}=
    x_0^2\left[-\delta_{{\lambda_{\rm E}},{ i}\nu}
+\delta_{{\lambda_{\rm E}},-{ i}\nu}\right] \label{eq:etr2} 
\end{alignat}
with $x_0=\sqrt{\hbar/2 M\nu}$. Using these results into
Eq.~(\ref{F}) we obtain Eqs.~(\ref{F:+}) and~(\ref{F:-}).

\section{Ion in the node of the standing wave}
\label{app:node}

In this appendix we evaluate explicitly the contribution $F_0(\lambda_{\rm I})$ and show that it identically vanish in the case where the atom is at the node of the standing wave. 

Let us first discuss the terms contributing to $F_0(\lambda_{\rm I})$. In the node of the standing wave $\rho_0=\ket g\bra g$. By using the relation $D_a\rho_0=0$ ($a=0,1,2=$), the terms which do not trivially vanish are 
\begin{equation}
\label{F:0:SW}
F_0(\lambda_I)=f_1^0(\lambda_{\rm I}) +f_5^0(\lambda_{\rm I}) +f_6^0(\lambda_{\rm I}) 
\end{equation}
where $f_j^0$ are the components of the terms 
\begin{eqnarray}
f_1(\lambda_{\rm I}) = {\rm Tr}\{D_0^{\dagger}{\cal P}^{\lambda}_{1}D_0\varrho_1\}\label{f:1}\\
f_5(\lambda_{\rm I}) = {\rm Tr}\{D_0^{\dagger}{\cal P}^{\lambda}_{0}D_0\varrho_2\}\label{f:5}\\
f_6(\lambda_{\rm I}) = {\rm Tr}\{D_0^{\dagger}{\cal P}^{\lambda}_{0}D_1\varrho_1\}\label{f:6}
\end{eqnarray}
which contribute to $F_0(\lambda_{\rm I})$. 
Their explicit form is found with the internal eigenelements, which at the node take the form
\begin{alignat}{2}
\rho^{\rm st}&=\ket g\bra g,& \check\rho^{\rm st}&=\uop_{\rm E},\nonumber\\
\rho^{0}&=\ket e\bra e-\ket g\bra g, \quad & \check\rho^{0}&=\ket e\bra e,\nonumber\\
\rho^{+}&=\ket e\bra g, &\check\rho^{+}&=\ket g\bra e\nonumber\\
\rho^{-}&=\ket g\bra e,& \check\rho^{-}&=\ket e\bra g\nonumber
\end{alignat}
with the corresponding eigenvalues $\lambda^{\rm st}=0$, $\lambda_{\rm I}^{0} = -\gamma$  and $\lambda_{\rm I}^{\pm} = -\gamma/2\pm i\delta$. Substituting into Eq.~(\ref{f:6}) one finds $f_6^0(\lambda)=0$. The other two terms  give at $\lambda_{\rm E}=0$
\begin{alignat}{1}
f_1^0(\lambda_{\rm I}) &= -\delta_{\lambda_{\rm I},\lambda^-} \frac{\Omega^2 k^2\cos^2\theta}{4}\sum_{\lambda_{\rm E}^\prime}\frac{\Tr_{\rm E}\{x \calU^{\lambda_{\rm E}^\prime}x\mu\}}{(\lambda^--\lambda_{\rm E}^\prime)(\lambda^++\lambda_{\rm E}^\prime)}\nonumber\\
f_5^0(\lambda_{\rm I}) &= \delta_{\lambda_{\rm I},\lambda^-} \frac{\Omega^2 k^2\cos^2\theta}{4}\times\nonumber\\
&\sum_{\lambda_{\rm E}^\prime}\Big[
\frac{\Tr_{\rm E}\{x \calU^{\lambda_{\rm E}^\prime}x\mu\}}{\lambda^0(\lambda^++\lambda_{\rm E}^\prime)}+\frac{\Tr_{\rm E}\{x \calU^{\lambda_{\rm E}^\prime}\mu x\}}{\lambda^0(\lambda^-+\lambda_{\rm E}^\prime)}
\Big]\nonumber.
\end{alignat}
Using the expressions for the external traces, Eqs.~(\ref{eq:etr1}) and (\ref{eq:etr2}), and the explicit form of eigenelements and eigenvectors, one finds that $f_1^0(\lambda_{\rm I})=-f_5^0(\lambda_{\rm I})$. Thus, the two terms mutually cancel in Eq.~(\ref{F:0:SW}) and hence $F_0(\lambda_{\rm I}) = 0$.
\end{appendix}

\end{document}